\documentclass[english]{article}
\usepackage[T1]{fontenc}
\usepackage[latin9]{inputenc}
\usepackage{geometry}
\usepackage{amsmath}
\usepackage{amssymb}

\makeatletter
\usepackage{indentfirst}

\makeatother

\usepackage{babel}
\begin{document}
\title{\textbf{\Large{}Wigner rotation and Euler angle parametrization }}
\author{{\normalsize{}Leehwa Yeh}\thanks{Electronic mail: yehleehwa@gmail.com}}
\date{\textit{\small{}Shing-Tung Yau Center, National Yang Ming Chiao Tung
University, Hsinchu 30010, Taiwan}}

\maketitle
\noindent Analogous to the famous Euler angle parametrization in three-dimensional
Euclidean space, a reflection-free Lorentz transformation in (2+1)-dimensional
Minkowski space can be decomposed into three simple parts. Applying
this decomposition to the Wigner rotation problem, we are able to
show the related mathematics becomes much simpler and the physical
meanings more comprehensible and enlightening. 

\section*{I. INTRODUCTION}

First discovered by Silberstein, then rediscovered by Thomas \cite{key-1,key-2},
the phenomenon that two successive non-parallel boosts (i.e., Lorentz
transformations that contain neither rotation nor reflection) lead
to a boost and a rotation is generally called the Wigner rotation
\cite{key-3}. It has been studied by many authors for almost a century
\cite{key-4,key-5,key-6,key-7,key-8,key-9}, yet the conclusion is
still ``paradoxical'' for most people. \textquotedblleft The spatial
rotation resulting from the successive application of two non-parallel
Lorentz transformations has been declared every bit as paradoxical
as the more frequently discussed apparent violations of common sense,
such as the so-called `twin paradox'...\textquotedblright{} said Goldstein
in his classic work \textit{Classical Mechanics} \cite{key-10}.

To explain how the puzzlement arises, we start with the classical
counterpart of a boost transformation, the so-called Galilean transformation
between two inertial frames with relative velocity $\vec{V}=[V_{x},V_{y},V_{z}]$,
\begin{align}
x' & =x-V_{x}t,\nonumber \\
y' & =y-V_{y}t,\nonumber \\
z' & =z-V_{z}t,\nonumber \\
t' & =t.\label{eq:01}
\end{align}
From a geometrical point of view, the first three lines of Eq. (\ref{eq:01})
constitute a time-dependent passive translation. It implies that a
Galilean transformation follows the rules of a three-dimensional translation,
e.g., the composition of two translations leads to another one, and
this net translation does not depend on the order in which the original
two are undertaken. Being the relativistic version of Galilean transformation,
a boost is often believed to possess this same property.

As a matter of fact, relativistic boosts do not commute unless their
directions are parallel, i.e., their composition is order-dependent
in general. Moreover, as mentioned earlier, the composition of two
non-parallel boosts is not a single boost, but a boost along with
a rotation. This is quite counter-intuitive since the rotation seems
to emerge out of nowhere from the three-dimensional point of view.
Getting to the bottom of the matter, it is inappropriate to analogize
boost to translation since the former is essentially a sort of rotation
(or more precisely a pseudo-rotation) in four-dimensional spacetime.
Therefore, the Wigner rotation may be regarded as a geometric problem
that involves both rotation and pseudo-rotation, and the mathematical
complexity is enough to cloud those subtle physical meanings. 

If the mathematics could be substantially simplified, we believe the
physical meanings of the Wigner rotation would become clear, and people
would find this phenomenon is not so counter-intuitive as usually
thought. To achieve this goal, we develop a formulation analogous
to the Euler angle parametrization of $SO(3)$, i.e., decomposing
a reflection-free Lorentz transformation in (2+1)-dimensional Minkowski
space into a product of two rotations and one pseudo-rotation (Sec.
II). As a demonstration of its effectiveness, we show how simple it
is to derive important rules about the Wigner rotation problem (Sec.
III) and how little mathematical knowledge is needed to calculate
the most general Wigner angle (Sec. IV). Physical insights into Wigner
rotation via this decomposition are discussed in Sec. III, and a rigorous
proof of this decomposition is provided in the Appendix. 

\section*{II. PRELIMINARIES}

\subsection*{A. (2+1)-dimensional Minkowski space}

Although physical spacetime is the (3+1)-dimensional Minkowski space
$\mathbb{R}^{3,1}$, physicists work on its subspace in many cases
without losing generality. For example, when discussing a pure (i.e.,
rotation-free and reflection-free) Lorentz transformation between
two inertial frames with the Minkowski coordinates $(x,y,z,ct)$ and
$(x',y',z',ct')$, where $c$ is the light speed, we may assume the
relative velocity $\vec{V}$ is along the $x$-direction and consider
just the transformation between $(x,ct)$ and $(x',ct')$, which is
represented by the formula
\begin{equation}
\begin{pmatrix}x'\\
ct'
\end{pmatrix}=\begin{pmatrix}\gamma & -\beta\gamma\\
-\beta\gamma & \gamma
\end{pmatrix}\begin{pmatrix}x\\
ct
\end{pmatrix}=\begin{pmatrix}\cosh\eta & -\sinh\eta\\
-\sinh\eta & \cosh\eta
\end{pmatrix}\begin{pmatrix}x\\
ct
\end{pmatrix},\label{eq:02}
\end{equation}
$\text{where }\beta=\tanh\eta=V/c\,\text{\,and }\gamma=\cosh\eta=1/\sqrt{1-\beta^{2}}.$
This transformation is usually called a boost along the $x$-direction
or an $x$-direction boost. From a geometric point of view, it is
a pseudo-rotation around $(0,0)$ in the (1+1)-dimensional Minkowski
space $\mathbb{R}^{1,1}.$ 

Similarly, since the Wigner rotation problem involves only two relative
velocities, it is legitimate to put them in the $xy$-plane so that
none of the $z$-components show up in the calculations. Therefore,
our discussion will be restricted to the (2+1)-dimensional Minkowski
space $\mathbb{R}^{2,1},$ which is sufficient for us to derive all
of the related results. 

It is apparent $x^{2}-c^{2}t^{2}$ is an invariant under the transformation
Eq. (\ref{eq:02}). When we use this invariant as the criterion for
the (1+1)-dimensional Lorentz transformation, the reflections such
as $x\longrightarrow-x$ or $ct\longrightarrow-ct$ will be included
as well. It is straightforward to generalize this criterion to (2+1)-dimensional
Minkowski space, i.e., we may define the Lorentz transformation in
this space as the one that preserves $x^{2}+y^{2}-c^{2}t^{2}$. Clearly,
both of the $x$-direction and $y$-direction boosts as well as the
$xy$-plane rotation are special cases of this (2+1)-dimensional Lorentz
transformation. 

There are many similarities between $\mathbb{R}^{2,1}$ and the three-dimensional
Euclidean space $\mathbb{R}^{3}$. For example, the three-dimensional
rotation by the angle $\theta$ about the axis $[-\sin\phi,\cos\phi,0]$
takes the form
\begin{equation}
\ensuremath{\begin{pmatrix}x'\\
y'\\
z'
\end{pmatrix}=\begin{pmatrix}\cos\phi & -\sin\phi & 0\\
\sin\phi & \cos\phi & 0\\
0 & 0 & 1
\end{pmatrix}\begin{pmatrix}\cos\theta & 0 & -\sin\theta\\
0 & 1 & 0\\
\sin\theta & 0 & \cos\theta
\end{pmatrix}\begin{pmatrix}\cos\phi & \sin\phi & 0\\
-\sin\phi & \cos\phi & 0\\
0 & 0 & 1
\end{pmatrix}\begin{pmatrix}x\\
y\\
z
\end{pmatrix},}\label{eq:03}
\end{equation}
while the transformation
\begin{equation}
\begin{pmatrix}x'\\
y'\\
ct'
\end{pmatrix}=\begin{pmatrix}\cos\phi & -\sin\phi & 0\\
\sin\phi & \cos\phi & 0\\
0 & 0 & 1
\end{pmatrix}\begin{pmatrix}\cosh\eta & 0 & -\sinh\eta\\
0 & 1 & 0\\
-\sinh\eta & 0 & \cosh\eta
\end{pmatrix}\begin{pmatrix}\cos\phi & \sin\phi & 0\\
-\sin\phi & \cos\phi & 0\\
0 & 0 & 1
\end{pmatrix}\begin{pmatrix}x\\
y\\
ct
\end{pmatrix}\label{eq:04}
\end{equation}
represents a pseudo-rotation around the corresponding axis $[-\sin\phi,\cos\phi,0]$
in $\mathbb{R}^{2,1}$, and it meets the criterion of being a (2+1)-dimensional
Lorentz transformation. However, note that the product of the three
transformation matrices in Eq. (\ref{eq:03}) is an orthogonal matrix,
i.e., its transpose is equal to its inverse, while that of Eq. (\ref{eq:04})
is a symmetric one, i.e., its transpose is itself.

Finally, if we express $x^{2}+y^{2}-c^{2}t^{2}$ as a matrix product
\begin{equation}
\begin{pmatrix}x & y & ct\end{pmatrix}\begin{pmatrix}1 & 0 & 0\\
0 & 1 & 0\\
0 & 0 & -1
\end{pmatrix}\begin{pmatrix}x\\
y\\
ct
\end{pmatrix}=:\begin{pmatrix}x & y & ct\end{pmatrix}\text{g}\begin{pmatrix}x\\
y\\
ct
\end{pmatrix},
\end{equation}
it is easy to see that the necessary and sufficient condition for
a $3\times3$ matrix $L$ to represent a (2+1)-dimensional Lorentz
transformation is $L^{\top}\text{g}L=\text{g}$, where the superscript
$\top$ represents taking transpose of the matrix. Accordingly, if
we have two (2+1)-dimensional Lorentz transformation matrices, say
$L_{1}$ and $L_{2}$, then both $L_{1}L_{2}$ and $L_{2}L_{1}$ are
also Lorentz transformations of this kind. (For the sake of brevity,
we will henceforward not distinguish between the transformation and
its matrix representation.)

\subsection*{B. Euler angles and their Minkowski counterparts}

The theory of Euler angles guarantees that any proper (i.e., reflection-free)
rotation in $\mathbb{R}^{3}$ can be decomposed into three simple
proper rotations with each of them keeping one coordinate axis fixed.
Among those commonly used Euler angle parametrizations, the one suits
us most is
\begin{equation}
\begin{pmatrix}x'\\
y'\\
z'
\end{pmatrix}=\begin{pmatrix}\cos\psi & -\sin\psi & 0\\
\sin\psi & \cos\psi & 0\\
0 & 0 & 1
\end{pmatrix}\begin{pmatrix}\cos\theta & 0 & -\sin\theta\\
0 & 1 & 0\\
\sin\theta & 0 & \cos\theta
\end{pmatrix}\begin{pmatrix}\cos\phi & \sin\phi & 0\\
-\sin\phi & \cos\phi & 0\\
0 & 0 & 1
\end{pmatrix}\begin{pmatrix}x\\
y\\
z
\end{pmatrix},\label{eq:06}
\end{equation}
where $0\le\phi,\psi<2\pi\text{ and }0\le\theta\le\pi$. 

Based on the analogy between Eqs. (\ref{eq:03}) and (\ref{eq:04}),
it is reasonable to assume that the (2+1)-dimensional counterpart
of Eq. (\ref{eq:06}) takes the form
\begin{equation}
\ensuremath{\begin{pmatrix}x'\\
y'\\
ct'
\end{pmatrix}=\begin{pmatrix}\cos\psi & -\sin\psi & 0\\
\sin\psi & \cos\psi & 0\\
0 & 0 & 1
\end{pmatrix}\begin{pmatrix}\cosh\eta & 0 & -\sinh\eta\\
0 & 1 & 0\\
-\sinh\eta & 0 & \cosh\eta
\end{pmatrix}\begin{pmatrix}\cos\phi & \sin\phi & 0\\
-\sin\phi & \cos\phi & 0\\
0 & 0 & 1
\end{pmatrix}\begin{pmatrix}x\\
y\\
ct
\end{pmatrix},}\label{eq:07}
\end{equation}
where $\eta\ge0$ and the ranges of $\phi$ and $\psi$ are the same
as those of Eq. (\ref{eq:06}). For an obvious reason, Eq. (\ref{eq:07})
will also be called an Euler decomposition in this paper. 

In the Appendix, we will prove that the (2+1)-dimensional reflection-free
Lorentz transformation 
\begin{equation}
\begin{pmatrix}x'\\
y'\\
ct'
\end{pmatrix}=\begin{pmatrix}\begin{array}{ccc}
L_{11} & L_{12} & L_{13}\\
L_{21} & L_{22} & L_{23}\\
L_{31} & L_{32} & L_{33}
\end{array}\end{pmatrix}\begin{pmatrix}x\\
y\\
ct
\end{pmatrix}
\end{equation}
can always be expressed as Eq. (\ref{eq:07}). If we further demand
$L_{33}\ne1$, then the decomposition is unique and the parameters
are determined by the following formulas: 
\begin{align}
 & \cos\phi=-L_{31}/\sqrt{L_{33}^{2}-1},\nonumber \\
 & \sin\phi=-L_{32}/\sqrt{L_{33}^{2}-1};\nonumber \\
 & \cosh\eta=L_{33};\nonumber \\
 & \cos\psi=-L_{13}/\sqrt{L_{33}^{2}-1},\nonumber \\
 & \sin\psi=-L_{23}/\sqrt{L_{33}^{2}-1}.\label{eq:09}
\end{align}

\subsection*{C. (2+1)-dimensional velocity}

A boost transformation takes place between two inertial frames; hence
each boost is defined by a constant velocity that is the relative
velocity between the frames. When a (2+1)-dimensional velocity undergoes
a boost $B(\vec{V})$ with $\vec{V}$ being the relative velocity,
the formula $W'=B(\vec{V})W$ is analogous to the boost transformation
of spacetime coordinates, where the column matrices $W$ and $W'$
represent the (2+1)-dimensional velocities in the old and the new
frames respectively. Conversely, the inverse boost transformation
$W=B(\vec{V}){}^{-1}W'$ allows us to calculate the (2+1)-dimensional
velocity in the old frame from that in the new one. 

Note that for an object with the spacetime coordinates $(x,y,ct)$,
its (2+1)-dimensional velocity $W$ is defined as
\begin{equation}
W^{\top}=\bigg(\frac{dx}{d\tau}\ \ \frac{dy}{d\tau}\ \ c\frac{dt}{d\tau}\bigg)=\frac{dt}{d\tau}\bigg(\frac{dx}{dt}\ \ \frac{dy}{dt}\ \ c\bigg)=:\gamma(\vec{v})\big(v_{x}\ \ v_{y}\ \ c\big),
\end{equation}
where $\tau$ is the proper time of this object, $\vec{v}=[v_{x},v_{y}]$
is the two-dimensional velocity with respect to the coordinate time,
and
\begin{equation}
\gamma(\vec{v})=\frac{1}{\sqrt{1-(v_{x}^{2}+v_{y}^{2})/c^{2}}}.
\end{equation}

Assume the object at rest in the new frame. Since its two-dimensional
velocity relative to the old frame equals the relative velocity $\vec{V}$
between those two frames, transforming its (2+1)-dimensional velocity
in the new frame back to that in the old reveals the information of
the boost velocity, 
\begin{align}
W & =B(\vec{V}){}^{-1}\begin{pmatrix}0\\
0\\
c
\end{pmatrix}=\gamma(\vec{V})\begin{pmatrix}V_{x}\\
V_{y}\\
c
\end{pmatrix}.
\end{align}

Taking Eq. (\ref{eq:04}) as an example, interpreting this pseudo-rotation
as a boost gives us the corresponding (2+1)-dimensional velocity 
\begin{align}
 & \begin{pmatrix}\cos\phi & -\sin\phi & 0\\
\sin\phi & \cos\phi & 0\\
0 & 0 & 1
\end{pmatrix}\begin{pmatrix}\cosh\eta & 0 & \sinh\eta\\
0 & 1 & 0\\
\sinh\eta & 0 & \cosh\eta
\end{pmatrix}\begin{pmatrix}\cos\phi & \sin\phi & 0\\
-\sin\phi & \cos\phi & 0\\
0 & 0 & 1
\end{pmatrix}\begin{pmatrix}0\\
0\\
c
\end{pmatrix}\nonumber \\
= & c\begin{pmatrix}\sinh\eta\cos\phi\\
\sinh\eta\sin\phi\\
\cosh\eta
\end{pmatrix}=\cosh\eta\begin{pmatrix}c\tanh\eta\cos\phi\\
c\tanh\eta\sin\phi\\
c
\end{pmatrix}.
\end{align}
This reveals that the boost velocity $\vec{V}=[V_{x},V_{y}]=c\tanh\eta[\cos\phi,\sin\phi]$.
Accordingly, Eq. (\ref{eq:04}) enumerates all possible boosts in
$\mathbb{R}^{2,1}$.

When a problem involves three inertial frames and two successive boosts,
say first $B(\vec{V}_{1})$ then $B(\vec{V}_{2})$, the (2+1)-dimensional
velocity of a rest object in the third frame can be transformed to
that in the first by
\begin{equation}
W=[B(\vec{V}_{2})B(\vec{V}_{1})]^{-1}\begin{pmatrix}0\\
0\\
c
\end{pmatrix}.\label{eq:14}
\end{equation}
The two-dimensional velocity contained in this (2+1)-dimensional velocity
is the composition of the two boost velocities in that order and can
be denoted by $\vec{V}_{1\oplus2}$. Therefore, we rewrite Eq. (\ref{eq:14})
as
\begin{equation}
W_{1\oplus2}=[B(\vec{V}_{2})B(\vec{V}_{1})]^{-1}\begin{pmatrix}0\\
0\\
c
\end{pmatrix}=\gamma(\vec{V}_{1\oplus2})\begin{pmatrix}(V_{1\oplus2})_{x}\\
(V_{1\oplus2})_{y}\\
c
\end{pmatrix}.\label{eq:15}
\end{equation}

\section*{III. WIGNER ROTATION}

\subsection*{A. Three rules}

By employing the Euler decomposition, we can derive three important
rules about the Wigner rotation problem systematically and effortlessly.
To be concise, the following shorthand notations will be used:
\begin{align}
\ensuremath{R(\phi)=\begin{pmatrix}\cos\phi & \sin\phi & 0\\
-\sin\phi & \cos\phi & 0\\
0 & 0 & 1
\end{pmatrix}} & \ensuremath{,}
\end{align}
\begin{equation}
B_{x}(-\eta)=\begin{pmatrix}\cosh\eta & 0 & -\sinh\eta\\
0 & 1 & 0\\
-\sinh\eta & 0 & \cosh\eta
\end{pmatrix},
\end{equation}
and 
\begin{equation}
B_{\phi}(-\eta)=R(-\phi)B_{x}(-\eta)R(\phi).\label{eq:18}
\end{equation}
Thereupon, Eqs. (\ref{eq:04}) and (\ref{eq:07}) can be respectively
expressed as
\begin{equation}
\begin{pmatrix}x'\\
y'\\
ct'
\end{pmatrix}=B_{\phi}(-\eta)\begin{pmatrix}x\\
y\\
ct
\end{pmatrix}\text{ and }\begin{pmatrix}x'\\
y'\\
ct'
\end{pmatrix}=R(-\psi)B_{x}(-\eta)R(\phi)\begin{pmatrix}x\\
y\\
ct
\end{pmatrix}.
\end{equation}

\textbf{Rule 1.} Two successive boosts are equivalent to a boost followed
or preceded by a rotation, which may be taken as the definition of
Wigner rotation.

Proof: As discussed in Sec. II A, for any two boosts $B_{1\text{ }}\text{and }B_{2}$
in $\mathbb{R}^{2,1},$ their product $B_{2}B_{1}$ must be a Lorentz
transformation in this space. Since these boosts contain no reflection
according to Eq. (\ref{eq:04}), neither does their product. Hence
we may apply the Euler decomposition to this product and demand that
\begin{equation}
B_{2}B_{1}=R(-\psi)B_{x}(-\eta)R(\phi).\label{eq:20}
\end{equation}
Note that the corresponding coordinate transformation is 
\begin{equation}
\begin{pmatrix}x'\\
y'\\
ct'
\end{pmatrix}=B_{2}B_{1}\begin{pmatrix}x\\
y\\
ct
\end{pmatrix},
\end{equation}
i.e., the boost $B_{1}$ takes place before $B_{2}$.

Referring to Eq. (\ref{eq:18}), Eq. (\ref{eq:20}) can be rearranged
in two different ways, 
\begin{align}
B_{2}B_{1} & =R(\phi-\psi)B_{\phi}(-\eta)=B_{\psi}(-\eta)R(\phi-\psi),\label{eq:22}
\end{align}
wherein the product $R(\phi-\psi)B_{\phi}(-\eta)$ corresponds to
a boost followed by a Wigner rotation, while $B_{\psi}(-\eta)R(\phi-\psi)$
to a boost preceded by the same rotation.

If the boosts $B_{1\text{ }}\text{and }B_{2}$ are parallel, i.e.,
their directions are the same or differ by $180^{\circ}$, then it
is easy to prove $\psi=\phi$ and $R(\phi-\psi)=I$. On the other
hand, $\psi\ne\phi$ implies $B_{1\text{ }}\text{and }B_{2}$ are
not parallel. In other words, the non-parallelism of boosts $B_{1\text{ }}\text{and }B_{2}$
is a necessary condition for the existence of a non-trivial Wigner
rotation. It is also a sufficient condition as will be proved in Sec.
IV B. 

\textbf{Rule 2.} If the order of the boosts in Rule 1 is exchanged,
then the sense of Wigner rotation is reversed.

Proof: Since the boost matrices are all symmetric and rotation matrices
all orthogonal, this rule can be proved by taking the transpose of
Eq. (\ref{eq:22}),
\begin{align}
B_{1}B_{2} & \ensuremath{=}B_{\phi}(-\eta)R(\psi-\phi)=R(\psi-\phi)B_{\psi}(-\eta).\label{eq:23}
\end{align}

For convenience sake, $\psi-\phi$ will be called Wigner angle from
now on.

\textbf{Rule 3.} The two-dimensional velocities corresponding to $B_{2}B_{1}$
and $B_{1}B_{2}$ in the previous rules differ only by a Wigner angle
\cite{key-8,key-9,key-20}.

Proof: According to Eq. (\ref{eq:15}), the two-dimensional velocities
$\vec{V}_{1\oplus2}$ and $\vec{V}_{2\oplus1}$ are generated by $B_{2}B_{1}$
and $B_{1}B_{2}$ respectively via the following formulas:
\begin{align}
W_{1\oplus2} & =(B_{2}B_{1})^{-1}\begin{pmatrix}0\\
0\\
c
\end{pmatrix};\label{eq:24}
\end{align}
\begin{equation}
W_{2\oplus1}=(B_{1}B_{2})^{-1}\begin{pmatrix}0\\
0\\
c
\end{pmatrix}.\label{eq:25}
\end{equation}
Substituting Eq. (\ref{eq:20}) into Eq. (\ref{eq:24}) gives us
\begin{equation}
W_{1\oplus2}=R(-\phi)B_{x}(\eta)R(\psi)\begin{pmatrix}0\\
0\\
c
\end{pmatrix}=c\begin{pmatrix}\sinh\eta\cos\phi\\
\sinh\eta\sin\phi\\
\cosh\eta
\end{pmatrix}.\label{eq:26}
\end{equation}
Similarly, substituting the transpose of Eq. (\ref{eq:20}) into Eq.
(\ref{eq:25}) leads to 
\begin{equation}
W_{2\oplus1}=R(-\psi)B_{x}(\eta)R(\phi)\begin{pmatrix}0\\
0\\
c
\end{pmatrix}=c\begin{pmatrix}\sinh\eta\cos\psi\\
\sinh\eta\sin\psi\\
\cosh\eta
\end{pmatrix}.\label{eq:27}
\end{equation}
Thus we find $\vec{V}_{1\oplus2}=c\tanh\eta[\cos\phi,\sin\phi]$ and
$\vec{V}_{2\oplus1}=c\tanh\eta[\cos\psi,\ensuremath{\sin}\psi]$,
i.e., their magnitudes are the same while their directions differ
by a Wigner angle $\psi-\phi$.

The physical meaning of Rule 3 is as follows: Consider three inertial
frames $K_{A},\ K_{B},$ and $K_{C}$. If $K_{B}$ results from boosting
$K_{A}$ by $B_{1}$ and $K_{C}$ from boosting $K_{B}$ by $B_{2},$
then a rest observer in $K_{A}$ finds the two-dimensional velocity
of $K_{C}$ is $c\tanh\eta[\cos\phi,\sin\phi]$. On the other hand,
if $K_{B}$ is from boosting $K_{A}$ by $B_{2}$ and $K_{C}$ from
boosting $K_{B}$ by $B_{1},$ the same observer will find $K_{C}$
moving with the same speed but toward the direction $[\cos\psi,\sin\psi]$.

\subsection*{B. Two kinds of Wigner rotations}

Assume the spacetime coordinates of an inertial frame $K$ are $(x,y,ct)$.
Applying Eq. (\ref{eq:23}) to this frame yields two equivalent results.
Both of them correspond to the same transformed frame $K'$ with the
coordinates $(x',y',ct')$,

\begin{equation}
\begin{pmatrix}x'\\
y'\\
ct'
\end{pmatrix}=B_{\phi}(-\eta)R(\psi-\phi)\begin{pmatrix}x\\
y\\
ct
\end{pmatrix}=R(\psi-\phi)B_{\psi}(-\eta)\begin{pmatrix}x\\
y\\
ct
\end{pmatrix}.\label{eq:28}
\end{equation}
Although being mathematically equivalent, they convey significantly
different physical meanings. According to a rest observer in the original
frame $K$, the first $R(\psi-\phi)$ operates on a rest frame, i.e.,
$K$ itself, while the second is responsible for rotating a moving
frame, which is $K$ boosted by $B_{\psi}(-\eta)$. 

In comparing them, the first rotation is more intriguing. This is
because the first equality in Eq. (\ref{eq:28}) is equivalent to
\begin{equation}
B_{\phi}(\eta)\begin{pmatrix}x'\\
y'\\
ct'
\end{pmatrix}=R(\psi-\phi)\begin{pmatrix}x\\
y\\
ct
\end{pmatrix},\label{eq:29}
\end{equation}
where $B_{\phi}(\eta)=R(-\phi)B_{x}(\eta)R(\phi)=R(-\phi-180^{\circ})B_{x}(-\eta)R(\phi+180^{\circ}).$ 

The physical meaning of Eq. (\ref{eq:29}) is the following. When
we apply $B_{\phi}(\eta)$ to $K'$ to obtain a rest frame, the new
frame will differ from the original rest frame by a Wigner angle $\psi-\phi$.
In other words, it is possible to engineer a Wigner rotation in two-dimensional
Euclidean space since the rotation of a rest frame does not involve
the temporal dimension \cite{key-20}. 

\section*{IV. WIGNER ANGLE}

In principle, for any two (2+1)-dimensional boosts $B_{1\text{ }}\text{and }B_{2}$,
we can always use Eq. (\ref{eq:09}) to calculate the corresponding
$(\phi,\eta,\psi)$ and obtain Wigner angle $\psi-\phi$ for the product
$B_{2}B_{1}$. In practice, there is an easier way as shown in the
following two examples. 

\subsection*{A. Perpendicular case}

If the directions of $B_{1}\text{ and }B_{2}$ are perpendicular to
each other, it is legitimate to specify 
\begin{equation}
\ensuremath{B_{1}=B_{x}(-\eta_{1})\text{ and}\ }\ensuremath{B_{2}=\begin{pmatrix}1 & 0 & 0\\
0 & \cosh\eta_{2} & -\sinh\eta_{2}\\
0 & -\sinh\eta_{2} & \cosh\eta_{2}
\end{pmatrix}}=:B_{y}(-\eta_{2}),
\end{equation}
where $\eta_{1}$ and $\eta_{2}$ are positive. As discussed in Sec.
III A, we may write
\begin{equation}
B_{2}B_{1}=R(-\psi_{p})B_{x}(-\eta_{p})R(\phi_{p}),
\end{equation}
where the subscript $p$ denotes ``perpendicular.'' 

Now instead of using Eq. (\ref{eq:09}) to express $(\phi_{p},\eta_{p},\psi_{p})$
in terms of those elements in $B_{2}B_{1}$, we calculate the following
(2+1)-dimensional velocities:
\begin{equation}
W_{1\oplus2}=(B_{2}B_{1})^{-1}\begin{pmatrix}0\\
0\\
c
\end{pmatrix}=B_{x}(\eta_{1})B_{y}(\eta_{2})\begin{pmatrix}0\\
0\\
c
\end{pmatrix}=c\begin{pmatrix}\sinh\eta_{1}\cosh\eta_{2}\\
\sinh\eta_{2}\\
\cosh\eta_{1}\cosh\eta_{2}
\end{pmatrix};
\end{equation}
\begin{equation}
W_{2\oplus1}=(B_{1}B_{2})^{-1}\begin{pmatrix}0\\
0\\
c
\end{pmatrix}=B_{y}(\eta_{2})B_{x}(\eta_{1})\begin{pmatrix}0\\
0\\
c
\end{pmatrix}=c\begin{pmatrix}\sinh\eta_{1}\\
\cosh\eta_{1}\sinh\eta_{2}\\
\cosh\eta_{1}\cosh\eta_{2}
\end{pmatrix}.
\end{equation}
Comparing these results with Eqs. (\ref{eq:26}) and (\ref{eq:27}),
we obtain what we are looking for, namely,
\begin{align}
 & \cos\phi{}_{p}=\sinh\eta_{1}\cosh\eta_{2}/\sinh\eta{}_{p},\nonumber \\
 & \sin\phi{}_{p}=\sinh\eta_{2}/\sinh\eta{}_{p};\nonumber \\
 & \cosh\eta_{p}=\cosh\eta_{1}\cosh\eta_{2};\nonumber \\
 & \cos\psi_{p}=\sinh\eta_{1}/\sinh\eta{}_{p},\nonumber \\
 & \sin\psi{}_{p}=\cosh\eta_{1}\sinh\eta_{2}/\sinh\eta{}_{p}.\label{eq:34}
\end{align}
The benefit of this method is there is no need to perform any $3\times3$
matrix multiplication. 

Once obtaining Eq. (\ref{eq:34}), with the help of the identity
\begin{equation}
\sinh^{2}\eta_{p}=(\cosh\eta_{1}\cosh\eta_{2}+1)(\cosh\eta_{1}\cosh\eta_{2}-1),
\end{equation}
we are able to calculate the sine and cosine functions of $\psi_{p}-\phi_{p}$,
\begin{equation}
\ensuremath{\sin(\psi_{p}-\phi_{p})}\ensuremath{=\frac{\sinh\eta_{1}\sinh\eta_{2}}{\cosh\eta_{1}\cosh\eta_{2}+1}}>0;\label{eq:36}
\end{equation}
\begin{equation}
\ensuremath{\cos(\psi_{p}-\phi_{p})=\frac{\cosh\eta_{1}+\cosh\eta_{2}}{\cosh\eta_{1}\cosh\eta_{2}+1}}>0,\label{eq:37}
\end{equation}
and find the range of $\psi_{p}-\phi_{p}$ is $(0,\pi/2)$.

If we adopt the $\beta\text{-\ensuremath{\gamma}}$ notation from
Sec. II A, 
\begin{equation}
\beta{}_{i}=\tanh\eta{}_{i}\text{ and }\gamma{}_{i}=1/\sqrt{1-\beta{}_{i}^{2}}=\cosh\eta{}_{i},
\end{equation}
the results in Eqs. (\ref{eq:36}) and (\ref{eq:37}) become
\begin{equation}
\ensuremath{\sin(\psi_{p}-\phi_{p})}\ensuremath{=\frac{\beta_{1}\beta_{2}\gamma_{1}\gamma_{2}}{\gamma_{1}\gamma_{2}+1}}\text{ and }\cos\ensuremath{(\psi_{p}-\phi_{p})=\frac{\gamma_{1}+\gamma_{2}}{\gamma_{1}\gamma_{2}+1},}
\end{equation}
and the results in Eq. (\ref{eq:34}) can be expressed elegantly via
the Euler decomposition,

\begin{equation}
B_{2}B_{1}=\begin{pmatrix}\frac{\beta_{1}}{\beta_{p}\gamma_{2}} & \frac{-\beta_{2}}{\beta_{p}} & 0\\
\\
\frac{\beta_{2}}{\beta_{p}} & \frac{\beta_{1}}{\beta_{p}\gamma_{2}} & 0\\
\\
0 & 0 & 1
\end{pmatrix}\begin{pmatrix}\gamma_{p} & 0 & -\beta_{p}\gamma_{p}\\
\\
0 & 1 & 0\\
\\
-\beta_{p}\gamma_{p} & 0 & \gamma_{p}
\end{pmatrix}\begin{pmatrix}\frac{\beta_{1}}{\beta_{p}} & \frac{\beta_{2}}{\beta_{p}\gamma_{1}} & 0\\
\\
\frac{-\beta_{2}}{\beta_{p}\gamma_{1}} & \frac{\beta_{1}}{\beta_{p}} & 0\\
\\
0 & 0 & 1
\end{pmatrix},
\end{equation}
where $\gamma_{p}=\gamma_{1}\gamma_{2},$ and 
\begin{equation}
\beta_{p}=\sqrt{1-\gamma_{p}^{-2}}=\sqrt{\beta_{1}^{2}+\beta_{2}^{2}-\beta_{1}^{2}\beta_{2}^{2}}.
\end{equation}

\subsection*{B. General case}

If the directions of the two boosts differ by an arbitrary angle $\Theta=2\theta\in[0,\pi]$,
the calculation of Wigner angle becomes complicated and is usually
performed using advanced mathematical tools \cite{key-6,key-7,key-8,key-20}.
With the aid of Euler decomposition, however, it is not necessary
to introduce any new tool and the derivation is just a little longer
than that of the perpendicular case.

In order to make the most of the symmetry, we consider the following
two boosts without losing generality:
\begin{equation}
\ensuremath{B_{1}=B_{-\theta}(-\eta_{1})=R(\theta)B_{x}(-\eta_{1})R(-\theta)};
\end{equation}
\begin{equation}
\ensuremath{B_{2}=B_{\theta}(-\eta_{2})=R(-\theta)B_{x}(-\eta_{2})R(\theta)},
\end{equation}
where $\eta_{1}$ and $\eta_{2}$ are positive. It is obvious that
their products are
\begin{align}
B_{2}B_{1} & =R(-\theta)B_{x}(-\eta_{2})R(2\theta)B_{x}(-\eta_{1})R(-\theta)
\end{align}
and
\begin{equation}
B_{1}B_{2}=R(\theta)B_{x}(-\eta_{1})R(-2\theta)B_{x}(-\eta_{2})R(\theta).
\end{equation}
We will denote the Euler decompositions of these two products by $R(-\psi)B_{x}(-\eta)R(\phi)$
and $R(-\phi)B_{x}(-\eta)R(\psi)$ respectively. There is no need
to provide a subscript for the parameters because the current case
is no less general than the one discussed in Sec. III A.

The perpendicular example suggests it might be wise to use the (2+1)-dimensional
velocities corresponding to $B_{2}B_{1}$ and $B_{1}B_{2}$ as shortcuts,
so our calculation begins by writing down the inverses of these two
products,
\begin{equation}
\ensuremath{(B_{2}B_{1})^{-1}=R(\theta)B_{x}(\eta_{1})R(-2\theta)B_{x}(\eta_{2})R(\theta)=R(-\phi)B_{x}(\eta)R(\psi);}\label{eq:46}
\end{equation}
\begin{align}
\ensuremath{(B_{1}B_{2})^{-1}=R(-\theta)B_{x}(\eta_{2})R(2\theta)B_{x}(\eta_{1})R(-\theta)} & =R(-\psi)B_{x}(\eta)R(\phi).\label{eq:47}
\end{align}
Combining Eqs. (\ref{eq:46}) and (\ref{eq:26}) leads to 

\begin{equation}
W_{1\oplus2}=R(\theta)B_{x}(\eta_{1})R(-2\theta)B_{x}(\eta_{2})R(\theta)\begin{pmatrix}0\\
0\\
c
\end{pmatrix}=c\begin{pmatrix}\sinh\eta\cos\phi\\
\sinh\eta\sin\phi\\
\cosh\eta
\end{pmatrix},\label{eq:48}
\end{equation}
which generates the relations
\begin{alignat}{1}
\sinh\eta\cos\phi & =\cos\theta(\cos2\theta\cosh\eta_{1}\sinh\eta_{2}+\sinh\eta_{1}\cosh\eta_{2})+\sin\theta\sin2\theta\sinh\eta_{2},\label{eq:49}\\
\sinh\eta\sin\phi & =-\sin\theta(\cos2\theta\cosh\eta_{1}\sinh\eta_{2}+\sinh\eta_{1}\cosh\eta_{2})+\cos\theta\sin2\theta\sinh\eta_{2},\label{eq:50}\\
\cosh\eta & =\cosh\eta_{1}\cosh\eta_{2}+\cos2\theta\sinh\eta_{1}\sinh\eta_{2}.\label{eq:51}
\end{alignat}
Similarly, Eqs. (\ref{eq:47}) and (\ref{eq:27}) together give us
\begin{equation}
W_{2\oplus1}=R(-\theta)B_{x}(\eta_{2})R(2\theta)B_{x}(\eta_{1})R(-\theta)\begin{pmatrix}0\\
0\\
c
\end{pmatrix}=c\begin{pmatrix}\sinh\eta\cos\psi\\
\sinh\eta\sin\psi\\
\cosh\eta
\end{pmatrix}.\label{eq:52}
\end{equation}
The similarity between Eqs. (\ref{eq:48}) and (\ref{eq:52}) enables
us to obtain the following two relations from Eqs. (\ref{eq:49})
and (\ref{eq:50}) by change of notation,
\begin{equation}
\ensuremath{\sinh\eta\cos\psi=\cos\theta(\cos2\theta\sinh\eta_{1}\cosh\eta_{2}+\cosh\eta_{1}\sinh\eta_{2})+\sin\theta\sin2\theta\sinh\eta_{1},}\label{eq:53}
\end{equation}
\begin{equation}
\ensuremath{\sinh\eta\sin\psi=\sin\theta(\cos2\theta\sinh\eta_{1}\cosh\eta_{2}+\cosh\eta_{1}\sinh\eta_{2})-\cos\theta\sin2\theta\sinh\eta_{1}.}\label{eq:54}
\end{equation}

Now we are equipped to find out $\sin(\psi-\phi)$ and $\cos(\psi-\phi)$,
but it is better to make a detour to calculate $\tan[(\psi-\phi)/2]$
first. This is because the identity

\begin{equation}
\ensuremath{\tan\big(\frac{\psi-\phi}{2}\big)=\frac{\sin\psi-\sin\phi}{\cos\psi+\cos\phi}}\label{eq:55}
\end{equation}
eliminates the common $\sinh\eta$ term when calculating the ratio,
and the calculation will be simpler.

Substituting Eqs. (\ref{eq:49}), (\ref{eq:50}), (\ref{eq:53}),
and (\ref{eq:54}) to the right hand side of Eq. (\ref{eq:55}) and
ignoring those $\sinh\eta$'s, we find the numerator is proportional
to
\begin{align}
 & \sin\theta(\cos2\theta+1)(\cosh\eta_{1}\sinh\eta_{2}+\sinh\eta_{1}\cosh\eta_{2})-\cos\theta\sin2\theta(\sinh\eta_{1}+\sinh\eta_{2})\nonumber \\
= & \cos\theta\sin2\theta(\cosh\eta_{1}\sinh\eta_{2}+\sinh\eta_{1}\cosh\eta_{2}-\sinh\eta_{1}-\sinh\eta_{2}),
\end{align}
and the denominator proportional to
\begin{align}
 & \cos\theta(\cos2\theta+1)(\cosh\eta_{1}\sinh\eta_{2}+\sinh\eta_{1}\cosh\eta_{2})+\sin\theta\sin2\theta(\sinh\eta_{1}+\sinh\eta_{2})\nonumber \\
= & \cos\theta\cos2\theta(\cosh\eta_{1}\sinh\eta_{2}+\sinh\eta_{1}\cosh\eta_{2}-\sinh\eta_{1}-\sinh\eta_{2})\nonumber \\
+ & \cos\theta(\cosh\eta_{1}\sinh\eta_{2}+\sinh\eta_{1}\cosh\eta_{2}+\sinh\eta_{1}+\sinh\eta_{2}).
\end{align}
Thus Eq. (\ref{eq:55}) can be expressed as
\begin{equation}
\ensuremath{\tan\big(\frac{\psi-\phi}{2}\big)=\frac{\sin\Theta}{\cos\Theta+X(\eta_{1},\eta_{2})}},\label{eq:58}
\end{equation}
where $\Theta=2\theta\in[0,\pi]$, and
\begin{align}
X(\eta_{1},\eta_{2}) & =\frac{(\cosh\eta_{1}+1)\sinh\eta_{2}+\sinh\eta_{1}(\cosh\eta_{2}+1)}{(\cosh\eta_{1}-1)\sinh\eta_{2}+\sinh\eta_{1}(\cosh\eta_{2}-1)}\nonumber \\
 & =\frac{\coth\dfrac{\eta_{1}}{2}+\coth\dfrac{\eta_{2}}{2}}{\tanh\dfrac{\eta_{1}}{2}+\tanh\dfrac{\eta_{2}}{2}}=\coth\frac{\eta_{1}}{2}\coth\frac{\eta_{2}}{2}.
\end{align}

Since both $\eta_{1}$ and $\eta_{2}$ are positive, $X$ is always
finite and Eq. (\ref{eq:58}) leads to the conclusion that $\psi-\phi=0$
implies $\Theta=0\text{ or }\pi$. Therefore, the non-parallelism
of $B_{1\text{ }}\text{and }B_{2}$ is a sufficient condition for
the existence of a non-zero Wigner angle. Once the range of $\Theta$
is restricted to $(0,\pi)$, we can deduce from $\infty>X>1$ that
$0<\psi-\phi<\Theta$. 

Next step is to use Eq. (\ref{eq:58}) to derive the sine and cosine
functions of the half angle,
\begin{equation}
\ensuremath{\sin\big(\frac{\psi-\phi}{2}\big)=\frac{\sin\Theta}{\sqrt{1+2X\cos\Theta+X^{2}}}=\frac{\sinh\eta_{1}\sinh\eta_{2}\sin\Theta}{\sqrt{2(\cosh\eta+1)(\cosh\eta_{1}+1)(\cosh\eta_{2}+1)}}};
\end{equation}

\begin{equation}
\ensuremath{\cos\big(\frac{\psi-\phi}{2}\big)=\frac{\cos\Theta+X}{\sqrt{1+2X\cos\Theta+X^{2}}}=\frac{\cosh\eta+\cosh\eta_{1}+\cosh\eta_{2}+1}{\sqrt{2(\cosh\eta+1)(\cosh\eta_{1}+1)(\cosh\eta_{2}+1)}}},
\end{equation}
where $\cosh\eta$ is given in Eq. (\ref{eq:51}), and
\begin{equation}
X=\frac{\cosh\eta_{1}+1}{\sinh\eta_{1}}\frac{\cosh\eta_{2}+1}{\sinh\eta_{2}}
\end{equation}
has been used in the derivation.

Switching to the $\beta\text{-\ensuremath{\gamma}}$ notation, we
obtain the following results \cite{key-8,key-11,key-12}:
\begin{alignat}{1}
\sin\big(\frac{\psi-\phi}{2}\big) & \ensuremath{=\frac{\beta_{1}\beta_{2}\gamma_{1}\gamma_{2}\sin\Theta}{\sqrt{2(\gamma+1)(\gamma_{1}+1)(\gamma_{2}+1)}}=\sqrt{\frac{(\gamma_{1}-1)(\gamma_{2}-1)}{2(\gamma+1)}}\sin\Theta},\\
\cos\big(\frac{\psi-\phi}{2}\big) & \ensuremath{=\frac{\gamma+\gamma_{1}+\gamma_{2}+1}{\sqrt{2(\gamma+1)(\gamma_{1}+1)(\gamma_{2}+1)}}};\\
\sin(\psi-\phi) & \ensuremath{=\frac{\beta_{1}\beta_{2}\gamma_{1}\gamma_{2}(\gamma+\gamma_{1}+\gamma_{2}+1)}{(\gamma+1)(\gamma_{1}+1)(\gamma_{2}+1)}\sin\Theta},\\
\cos(\psi-\phi) & \ensuremath{=}1-\frac{(\gamma_{1}-1)(\gamma_{2}-1)}{\gamma+1}\sin^{2}\Theta=\frac{(\gamma+\gamma_{1}+\gamma_{2}+1)^{2}}{(\gamma+1)(\gamma_{1}+1)(\gamma_{2}+1)}-1,
\end{alignat}
where $\gamma=\gamma_{1}\gamma_{2}(1+\beta_{1}\beta_{2}\cos\Theta)$. 

\section*{V. CONCLUSION}

The Euler decomposition introduced in this paper is the most natural
and powerful tool for studying the Wigner rotation problem. Once the
mathematics is substantially simplified, the physical meanings are
easier to comprehend even for the beginners. 

\section*{APPENDIX: VALIDITY OF EULER DECOMPOSITION }

In this appendix, we provide a rigorous proof for the validity of
Euler decomposition of the reflection-free Lorentz transformation
in the (2+1)-dimensional Minkowski space $\mathbb{R}^{2,1}$. Our
proof begins with considering a general Lorentz transformation in
this space, 

\begin{equation}
L=\begin{pmatrix}\begin{array}{ccc}
L_{11} & L_{12} & L_{13}\\
L_{21} & L_{22} & L_{23}\\
L_{31} & L_{32} & L_{33}
\end{array}\end{pmatrix}.\tag{A1}\label{eq:67}
\end{equation}

From the discussion in Sec. II A, we know the matrix $L$ is neither
symmetric nor orthogonal in general. Instead, it satisfies the condition
$L^{\top}\text{g}L=\text{g}$, which implies $\det(L)=\pm1$ and $L$
is invertible. This condition and its equivalent $L\text{g}L^{\top}=\text{g}$
together generate twelve relations among the elements in $L$ (not
totally independent of course), wherein the following four will be
used in our proof:
\begin{alignat}{1}
L_{13}^{2}+L_{23}^{2} & =L_{33}^{2}-1,\tag{A2}\label{eq:68}\\
L_{31}^{2}+L_{32}^{2} & =L_{33}^{2}-1,\tag{A3}\label{eq:69}\\
L_{11}L_{31}+L_{12}L_{32} & =L_{13}L_{33},\tag{A4}\label{eq:70}\\
L_{21}L_{31}+L_{22}L_{32} & =L_{23}L_{33}.\tag{A5}\label{eq:71}
\end{alignat}

To exclude reflections from the transformation Eq. (\ref{eq:67}),
we first rule out those with $\det(L)=-1$, e.g.,
\begin{equation}
\ensuremath{\begin{pmatrix}-1 & 0 & 0\\
0 & 1 & 0\\
0 & 0 & 1
\end{pmatrix}\text{,\  }\begin{pmatrix}\cosh\eta & 0 & -\sinh\eta\\
0 & -1 & 0\\
-\sinh\eta & 0 & \cosh\eta
\end{pmatrix},}\text{ and }\begin{pmatrix}\cos\theta & \sin\theta & 0\\
-\sin\theta & \cos\theta & 0\\
0 & 0 & -1
\end{pmatrix}.\tag{A6}
\end{equation}
Then among the $L$'s with positive determinant, we have to rule out
those with negative $L_{33}$, which correspond to the transformations
containing both spatial and temporal reflections, e.g.,
\begin{equation}
\begin{pmatrix}-1 & 0 & 0\\
0 & 1 & 0\\
0 & 0 & -1
\end{pmatrix},\ \begin{pmatrix}-\cosh\eta & 0 & \sinh\eta\\
0 & 1 & 0\\
\sinh\eta & 0 & -\cosh\eta
\end{pmatrix},\text{ and }\begin{pmatrix}\cos\theta & \sin\theta & 0\\
\sin\theta & -\cos\theta & 0\\
0 & 0 & -1
\end{pmatrix}.\tag{A7}
\end{equation}
Since $L_{33}^{2}\ge1$ according to Eq. (\ref{eq:68}) or (\ref{eq:69}),
once the negative part is ruled out, the range of $L_{33}$ is reduced
to $[1,\infty)$. 

In summary, for Eq. (\ref{eq:67}) to contain no reflection, the matrix
$L$ has to satisfy: (i) $\det(L)=1$ and (ii) $L_{33}\ge1$ \cite{key-21}.

Equations (\ref{eq:68}) and (\ref{eq:69}) also tell us that $L_{33}=1$
implies $L_{13}=L_{23}=L_{31}=L_{32}=0$, and accordingly $L$ degenerates
to a rotation which corresponds to a special case of Eq. (\ref{eq:07}),
\begin{equation}
\begin{pmatrix}x'\\
y'\\
ct'
\end{pmatrix}=\begin{pmatrix}\cos(\psi-\phi) & -\sin(\psi-\phi) & 0\\
\sin(\psi-\phi) & \cos(\psi-\phi) & 0\\
0 & 0 & 1
\end{pmatrix}\begin{pmatrix}x\\
y\\
ct
\end{pmatrix}.\tag{A8}
\end{equation}
It is obvious that the Euler decomposition is valid for this degenerate
case, but with the result that the parameters $\phi\text{ and }\psi$
are not unique.

After we exclude all reflections and the degenerate case, a one-to-one
correspondence can be built between Eq. (\ref{eq:67}) and the transformation
matrix in Eq. (\ref{eq:07}) 
\begin{equation}
\begin{pmatrix}\cosh\eta\cos\psi\cos\phi+\sin\psi\sin\phi & \cosh\eta\cos\psi\sin\phi-\sin\psi\cos\phi & -\sinh\eta\cos\psi\\
\cosh\eta\sin\psi\cos\phi-\cos\psi\sin\phi & \cosh\eta\sin\psi\sin\phi+\cos\psi\cos\phi & -\sinh\eta\sin\psi\\
-\sinh\eta\cos\phi & -\sinh\eta\sin\phi & \cosh\eta
\end{pmatrix}.\tag{A9}
\end{equation}

This correspondence may be achieved through the following three steps. 

(1) Since $L_{33}>1$, we can always find a unique positive $\eta$
such that $L_{33}=\cosh\eta$, and accordingly $\sinh\eta=\sqrt{L_{33}^{2}-1}>0$.

(2) Since Eq. (\ref{eq:68}) is the same as $(-L_{13})^{2}+(-L_{23})^{2}=L_{33}^{2}-1$,
a unique $\psi\in[0,2\pi)$ exists such that
\begin{align}
 & -L_{13}=\sqrt{L_{33}^{2}-1}\cos\psi=\sinh\eta\cos\psi;\tag{A10}\\
 & -L_{23}=\sqrt{L_{33}^{2}-1}\sin\psi=\sinh\eta\sin\psi.\tag{A11}
\end{align}
For the same reason, Eq. (\ref{eq:69}) guarantees there is a unique
$\phi\in[0,2\pi)$ such that
\begin{align}
 & -L_{31}=\sqrt{L_{33}^{2}-1}\cos\phi=\sinh\eta\cos\phi;\tag{A12}\\
 & -L_{32}=\sqrt{L_{33}^{2}-1}\sin\phi=\sinh\eta\sin\phi.\tag{A13}
\end{align}

(3) So far we have obtained Eq. (\ref{eq:09}) which allows us to
determine a unique set $(\phi,\eta,\psi)$ by the matrix elements
$L_{13},L_{23},L_{31},L_{32}$, and $L_{33}$. What remains is to
verify that Eq. (\ref{eq:09}) is consistent with the other four relations:
\[
L_{11}=\cosh\eta\cos\psi\cos\phi+\sin\psi\sin\phi,
\]
\[
L_{12}=\cosh\eta\cos\psi\sin\phi-\sin\psi\cos\phi,
\]
\[
L_{21}=\cosh\eta\sin\psi\cos\phi-\cos\psi\sin\phi,
\]
\begin{equation}
L_{22}=\cosh\eta\sin\psi\sin\phi+\cos\psi\cos\phi.\tag{A14}
\end{equation}
This is equivalent to proving the following four equalities:
\[
L_{11}=(L_{33}L_{13}L_{31}+L_{23}L_{32})/(L_{33}^{2}-1),
\]
\[
L_{12}=(L_{33}L_{13}L_{32}-L_{23}L_{31})/(L_{33}^{2}-1),
\]
\[
L_{21}=(L_{33}L_{23}L_{31}-L_{13}L_{32})/(L_{33}^{2}-1),
\]
\begin{equation}
L_{22}=(L_{33}L_{23}L_{32}+L_{13}L_{31})/(L_{33}^{2}-1).\tag{A15}
\end{equation}
Taking Eqs. (\ref{eq:69})$-$(\ref{eq:71}) into account, these four
equalities can be reduced to the following two:
\begin{equation}
L_{21}L_{32}-L_{22}L_{31}=-L_{13},\tag{A16}\label{eq:82}
\end{equation}
\begin{equation}
L_{31}L_{12}-L_{32}L_{11}=-L_{23}.\tag{A17}\label{eq:83}
\end{equation}

In order to prove Eqs. (\ref{eq:82}) and (\ref{eq:83}), we express
the determinant of $L$ as (adopting the summation convention)
\begin{equation}
\det(L)=\frac{1}{3!}\epsilon_{ijk}\epsilon_{pqr}L_{ip}L_{jq}L_{kr}=1,\tag{A18}
\end{equation}
then transpose one of the two Levi-Civita symbols to the other side
of the equal sign \cite{key-13},
\begin{equation}
\epsilon_{pqr}L_{ip}L_{jq}L_{kr}=\epsilon_{ijk}.\tag{A19}
\end{equation}
Since $L^{\top}\text{g}L=\text{g}$ is equivalent to $\text{g}L\text{g}=(L^{\top})^{-1}$,
the above equality can be transformed to 
\begin{equation}
\epsilon_{pqr}L_{ip}L_{jq}=\epsilon_{ijk}(\text{g}L\text{g})_{kr}.\tag{A20}\label{eq:86}
\end{equation}
Both Eqs. (\ref{eq:82}) and (\ref{eq:83}) are special cases of Eq.
(\ref{eq:86}), the former corresponds to $(r,i,j)=(3,2,3)$ and the
latter to $(r,i,j)=(3,3,1).$ Q.E.D.

\section*{ACKNOWLEDGMENTS}

This paper is in memory of my father Colonel Li-jung Yeh (1930-2021)
and my thesis advisor Professor Geoffrey F. Chew (1924-2019).

\end{document}